\documentclass[conference]{IEEEtran}

\makeatletter
\def\ps@headings{%
\def\@oddhead{\mbox{}\scriptsize\rightmark \hfil \thepage}%
\def\@evenhead{\scriptsize\thepage \hfil \leftmark\mbox{}}%
\def\@oddfoot{}%
\def\@evenfoot{}}
\makeatother
\usepackage{amsmath}
\usepackage{amssymb}
\usepackage{graphicx}
\usepackage{subfigure}
\usepackage[lined,boxed,commentsnumbered, ruled]{algorithm2e}  
\usepackage{amsthm}
\usepackage{mathptm}
\usepackage{cite}
\usepackage{stfloats}
\usepackage{enumerate}

\begin{document}

\title{Triggercast: Enabling Wireless Collisions Constructive}

\author{\IEEEauthorblockN{Yin Wang$^{\dag\S}$, Yuan He$^{\dag\S}$, Dapeng Cheng$^{\dag}$, Yunhao Liu$^{\dag\S}$, Xiang-yang Li$^{\ddag\dag}$}
\IEEEauthorblockA{$^{\dag}$MOE Key Lab for Information System Security, School of Software, TNLIST,  Tsinghua University\\
$^{\S}$Department of Computer Science and Engineering, HKUST\\
$^{\ddag}$Department of Computer Science, Illinois Institute of Technology\\
\{wangyin00,chengdapeng\}@gmail.com, \{he, yunhao\}@greenorbs.com, xli@cs.iit.edu
}}

\maketitle

%
%
%
\begin{abstract}
It is generally considered that concurrent transmissions should be avoided in order to reduce collisions in wireless sensor networks.
Constructive interference (CI) envisions concurrent transmissions to \emph{positively} interfere at the receiver.
CI potentially allows orders of magnitude reductions in energy consumptions and improvements on link quality.
In this paper, we theoretically introduce a sufficient condition to construct CI with IEEE 802.15.4 radio for \emph{the first time}.
Moreover, we propose Triggercast, a distributed middleware, and show it is feasible to generate CI in TMote Sky sensor nodes.
To synchronize transmissions of multiple senders at the chip level, Triggercast effectively compensates propagation and radio processing delays, and has $95^{th}$ percentile synchronization errors of at most 250ns.
Triggercast also intelligently decides which co-senders to participate in simultaneous
transmissions, and aligns their transmission time to maximize the overall link PRR, under the condition of maximal system robustness.
Extensive experiments in real testbeds reveal that Triggercast significantly improves PRR from 5\% to 70\% with 7 concurrent senders.
We also demonstrate that Triggercast provides on average $1.3\times$ PRR performance gains, when integrated with existing data forwarding protocols.
\end{abstract}
\section{Introduction}
\begin{figure}
\centering
    \subfigure[Non-destructive interference]{\includegraphics[width=1.4in]{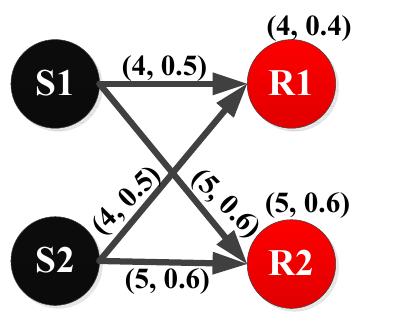}}
    \subfigure[Constructive interference]    {\includegraphics[width=1.2in]{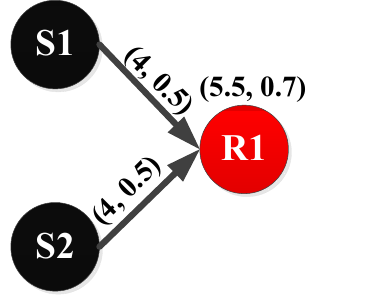}}
\caption{Both NDI and CI enable concurrency. Only CI improves RSSI and PRR. Here, we use (a, b) to describe a link, while a and b represent the RSSI and PRR respectively.}
{\label{Fig_NDIandCI}}
\end{figure}

\begin{figure}
\centering
    \subfigure[sender-initiated]{\includegraphics[width=1.6in]{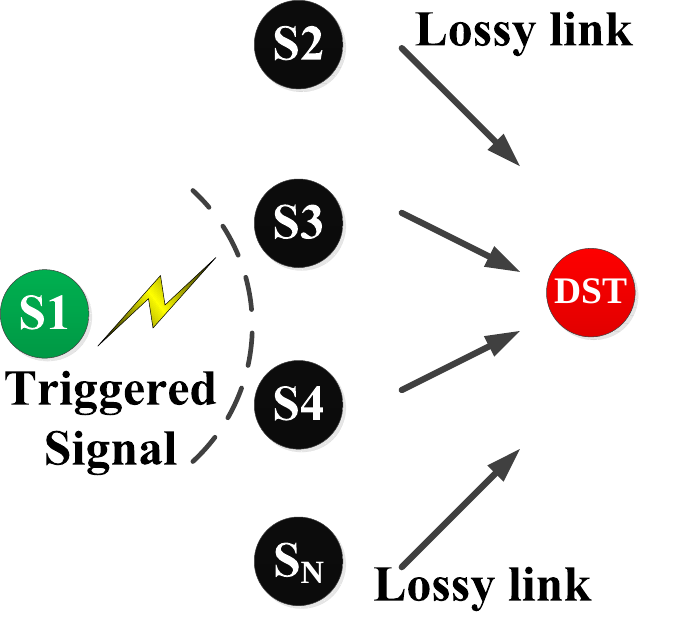}}
    \subfigure[receiver-initiated]    {\includegraphics[width=1.6in]{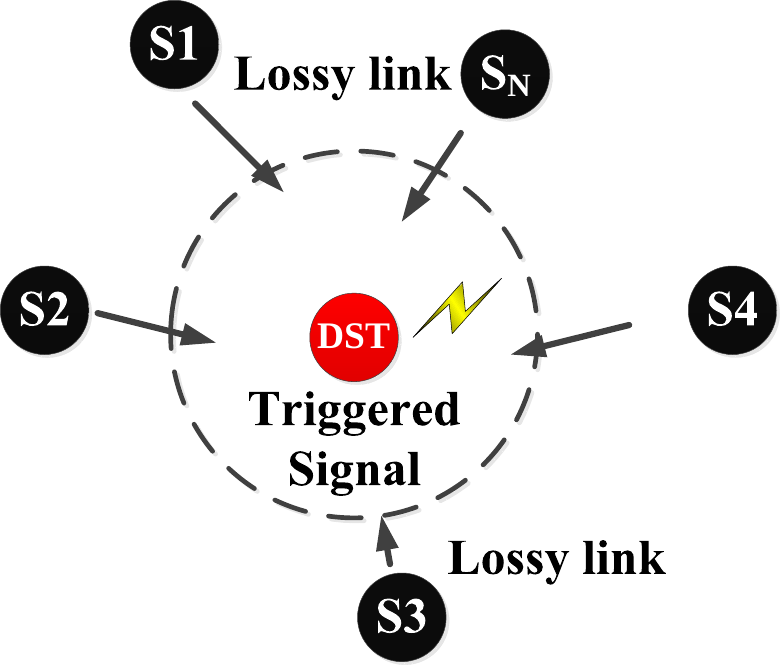}}
\caption{Triggercast: a radio triggered concurrent transmission architecture.}
{\label{Fig_TriggerCastArchitecture}}
\end{figure}

\begin{figure}
\centering
    \subfigure[Reduce transfer latency]{\includegraphics[width=1.5in]{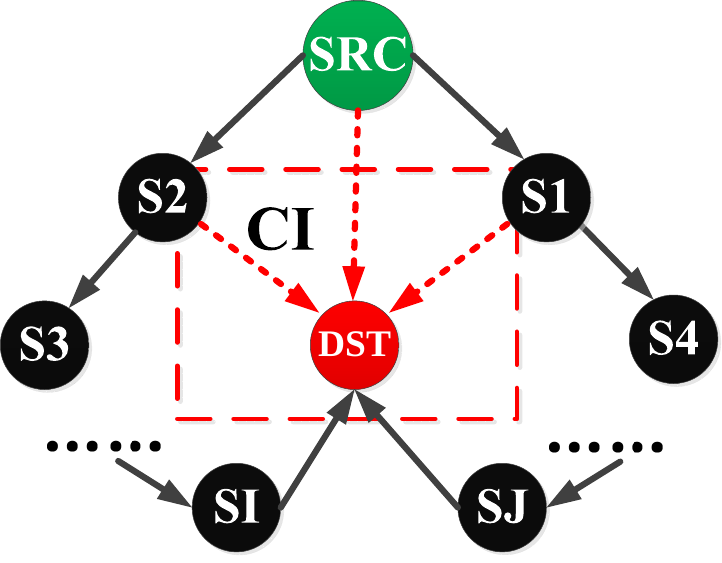}}
    \subfigure[Reduce data retransmissions]
    {\includegraphics[width=1.7in]{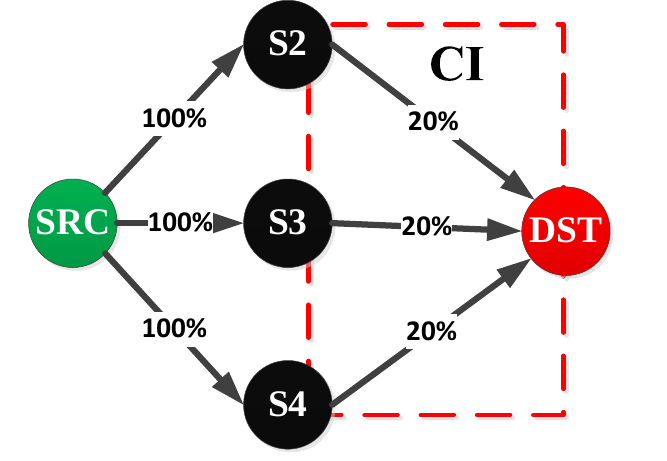}}
\caption{(a) Triggercast generates a new link from three disconnected links and thus reduce data forwarding latency. (b) Triggercast makes use of signal superposition to improve PRR and hence reduce retransmission times.}
{\label{Fig_OpportunisticRouting}}
\end{figure}

In wireless sensor networks (WSNs), it is widely accepted that simultaneous transmissions will result in packet collisions.
Recently, Backcast\cite{DuttaSensys10Backcast} and Glossy\cite{ferrari11Glossy} demonstrate that it is feasible for a common receiver to decode concurrent transmissions of an identical packet with high probability, if multiple transmissions are accurately synchronized.
Their works enable simultaneous transmissions to interfere non-destructively, namely to generate non-destructive interference (NDI), in order to enhance network concurrency.
By leveraging NDI, Glossy achieves nearly optimal network flooding latency.
Since NDI requires different nodes transmitting the same packet and thus may consume more power and require collaboration, one question naturally arises: is NDI constructive? Unfortunately, our extensive experiments disclose that NDI provides no guarantee of power gains and PRR improvements compared with the single best link (Fig. \ref{Fig_NDIandCI}(a)).\\
\indent  Our work aims to implement constructive interference (CI) in WSNs.
%
%
%
%
%
CI is especially attractive for WSNs, because it potentially improves energy efficiency, and thus mitigates the limited power supply issue.
A set of $N$ nodes can achieve an $N^2$-fold increase in the received power of \emph{baseband} signals, compared to a single node transmitting individually.
It indicates that, to achieve the same SNR, each node can reduce signal power with a factor of $\frac{1}{N^2}$, and the total power consumed by $N$ nodes can be $\frac{1}{N}$ of the transmitting power required by a single sender.
Moreover, simultaneously forwarding a packet can harness signal superposition gain, to improve RSSI and PRR (Fig. \ref{Fig_NDIandCI}(b)).\\
\indent However, implementing CI in WSNs is challenging due to the following reasons. First, simultaneous transmissions must be synchronized at the chip level, namely 0.5$\mu$s for IEEE 802.15.4 radio.
To generate NDI, Glossy's synchronization is sufficient, since it compensates most factors, such as clock drifts, software routine uncertainties of OS as well as asynchronous clocks (e.g., transmitter's radio and receiver's radio, MCU and radio module).
However, it is not sufficient to construct CI.
Experiments reveal that propagation delays and radio processing delays significantly influence CI generation.
Even worse, estimating radio processing delays is an especially challenging task, as it varies from packet to packet, depends on the SNR, and is affected by multi-path characteristics of the channel.
Besides, in the absence of a central controller or a shared clock (e.g., GPS), they can only rely on their own radio signal as a reference.\\
\indent Second, even if simultaneous transmissions are perfectly synchronized, i.e. no phase offset, they might not guarantee CI. The reason is because a radio signal has noise. Although signals are exactly aligned, noises also superpose. Whether SNR of the combined signal increases depends on SNRs and Tx powers of individual signals.\\
\indent Third, sensor nodes are always battery-powered, and have limited computational resources. It is difficult or even impossible to deploy complex signal processing algorithms in commercial of the shelf (COTS) sensor platforms.\\
\indent We propose Triggercast, a practical distributed middleware to generate CI in WSNs.
Triggercast enables a co-sender (sender-initiated Triggercast, Fig. \ref{Fig_TriggerCastArchitecture}(a)) or a receiver (receiver-initiated Triggercast, Fig. \ref{Fig_TriggerCastArchitecture}(b)) to trigger a radio signal, which acts as a common reference for all concurrent senders to implement synchronized transmissions.
The chip level synchronization (CLS) algorithm of Triggercast enables concurrent transmissions to be synchronized in 0.5$\mu$s, by compensating propagation and radio processing delays.
Our experiments demonstrate that CLS has $95^{th}$ percentile synchronization errors of at most 250ns.
The accuracy is bounded by the running frequency (4,194,304Hz) of on-board MCU of TMote Sky sensor node.
Triggercast's link selection and alignment (LSA) algorithm intelligently decides which co-senders to participate in simultaneous
transmissions, and aligns their transmission time to maximize the overall link PRR under the condition of maximal system robustness.
The underling CLS and LSA algorithms together ensure Triggercast to generate CI in a practical testbed.
Extensive experiments show that Triggercast can improve PRR from 5\% to 70\% with 7 senders, and from 50\% to 98.3\% with 6 senders.
Experiments also demonstrate that Triggercast on average brings a $1.3\times$ PRR performance gain of data forwarding in realistic deployments.\\
\indent Experimental results indicate that Triggercast can control topology without increasing Tx power or adding new nodes, which opportunistically reduces latency of data forwarding (in Fig. \ref{Fig_OpportunisticRouting}(a)).
Triggercast can also reduce packet retransmissions by improving PRR (in Fig. \ref{Fig_OpportunisticRouting}(b)).
For example, in Fig. \ref{Fig_OpportunisticRouting}(b), the ETX of traditional routing is $\text{ETX}_1={\raise0.7ex\hbox{$1$} \!\mathord{\left/
 {\vphantom {1 {0.2}}}\right.\kern-\nulldelimiterspace}
\!\lower0.7ex\hbox{${0.2}$}} + 1 = 6$,
while the ETX of Triggercast might be ${\raise0.7ex\hbox{$1$} \!\mathord{\left/
 {\vphantom {1 {0.3}}}\right.\kern-\nulldelimiterspace}
\!\lower0.7ex\hbox{${0.32}$}} + 1 \approx 4$ (the number 0.32 comes from real measurements).\\
\indent The contributions of this paper are summarized as follows.
\begin{enumerate}
\setlength{\itemsep}{-0.245ex}
\item
We are the \emph{first} to provide a \emph{theoretical sufficient condition} for generating CI in WSNs.
\item
We propose Triggercast, a practical middleware to ensure concurrent transmissions to interfere constructively.
The underlining CLS algorithm effectively evaluates and compensates propagation and radio processing delays.
\item
We implement Triggercast in real testbeds. Extensive experiments show Triggercast can construct CI in TMote Sky platforms. We integrate Triggercast into data forwarding protocols and show its performance gains.
\end{enumerate}
\section{Related Work}
\label{Sec_Relate}
\indent
Exploiting concurrent transmissions while suppressing interference is a promising direction, for its ability to decode packets from collisions, increase network throughput \cite{wang2010multicast,Capacity-INFOCOM12,chu2010opportunistic}, alleviate the broadcast storm problem of ackowledgements\cite{DuttaSensys10Backcast}, enhance packet transmission reliability, and reduce flooding latency \cite{ferrari11Glossy}.
Prior works can be categorized as signal processing based and physical-layer phenomenon based.\\
\indent Works based on signal processing include ANC \cite{katti2007embracing} for network coding, SIC \cite{halperin2008taking} and Zigzag \cite{gollakota2008zigzag} for interference cancelation, 802.11n+ \cite{lin2011random} for interference alignment in MIMO, AutoMAC \cite{Gudipati2012AutoMAC} for rateless coding, and full-duplex wireless radios \cite{jain2011practical}.
Those works leverage powerful software-defined radio platforms (e.g., USRP), and mainly aim at improving throughput in wireless networks.
Unfortunately, these signal processing algorithms can not be directly applied in WSNs, in which sensor nodes have insufficient computation resources and limited energy supplies.\\
\indent Physical-layer phenomenon based works mainly focus on exploring wireless radio properties of COTS transceivers.
Such physical-layer phenomena mainly include capture effect \cite{leentvaar1976captureeffect} and message-in-message (MIM) \cite{santhapuri2008message}.
Capture effect requires signal of interest is sufficient stronger than the sum of interference.
MIM needs special hardware support to continuously synchronize with the preamble of stronger signal.
Both capture effect and MIM can only decode the stronger signal at the cost of dropping the other signals.\\
\indent Recently, Backcast \cite{DuttaSensys10Backcast} experimentally discovers that, concurrent transmissions of short acknowledgment packets automatically generated by the radio hardware can interfere non-destructively.
This characteristic can be utilized to alleviate the ACK implosion problem \cite{ni1999broadcaststorm}.
Glossy \cite{ferrari11Glossy} advances this work of NDI by implementing designs such as interrupt compensation and precise timing controls.
Although multiple senders transmiting the same packet many times means consuming more power and needs cooperation, NDI is reasonable because it greatly reduces the time incurred by collision scheduling, and thus improves network throughput. The main purpose of our work is to make those wireless collisions interfere constructively. \\
\indent Triggercast's radio-triggered synchronization mechanism is comparable with those in SourceSync \cite{rahul2010sourcesync} and Glossy \cite{ferrari11Glossy}.
SourceSync exploits the fundamental property of FFTs to evaluate the radio processing delay, which varies
dynamically in multi-path channels.
Unfortunately, single-carrier communication (e.g., IEEE 802.15.4) systems cannot benefit from this method introduced by SourceSync.
Glossy is able to synchronize packet transmissions at the magnitude of sub-microseconds, which previously is considered too challenging to implement on COTS sensor platforms.
However, for signals to superpose constructively, we need more accurate synchronization algorithms to
compensate propagation delays and radio processing delays.
Triggercast aims to address this problem. \\

\section{Packet Transmissions Over Interferences}
\subsection{Background}
\begin{figure}
\centering
\includegraphics[width=3.5in]{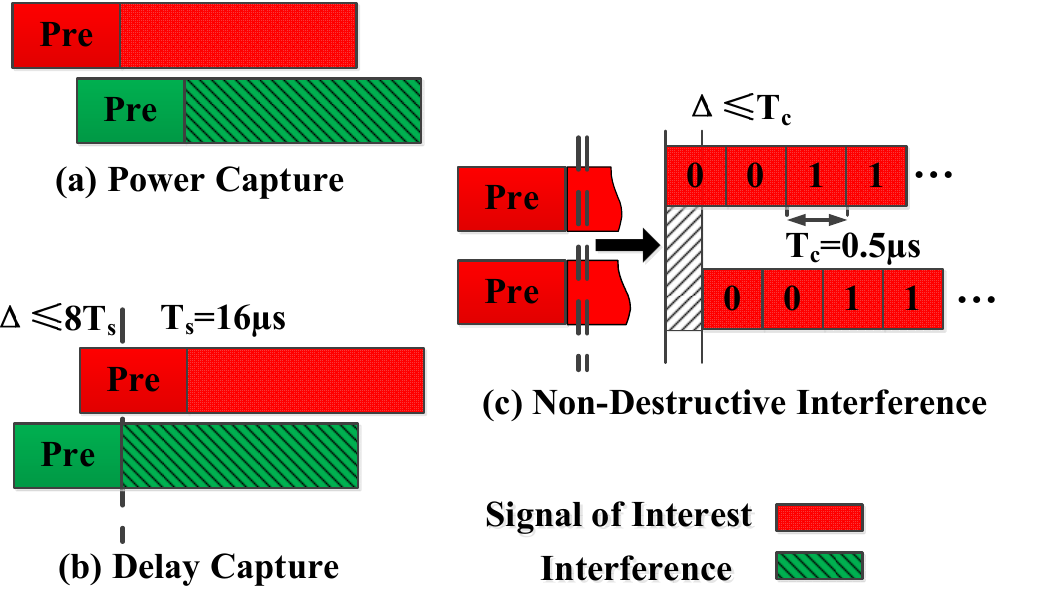}
\caption{Concurrent transmission techniques supported by COTS IEEE 802.15.4 transceivers.}
\label{Fig_ConCurrentOverview}
\end{figure}

\textbf{Concurrent Transmissions:} packet transmissions over interferences have been studied extensively as they can be utilized to help receivers to decode packets from collisions.
In WSNs, due to the limitations of low-cost sensor nodes and COTS radio transceivers (e.g., CC2420, AT86RF230), it is usually difficult or even impossible to leverage techniques such as network coding, successive interference cancelation or MIMO to accomplish packet transmissions while suppressing interference.
As a result, previous studies in WSNs mainly focus on exploring physical layer phenomena to realize concurrent packet transmissions.\\
\indent There are two well-known concurrent transmission techniques, namely capture effect \cite{leentvaar1976captureeffect} and MIM \cite{santhapuri2008message}.
It can be seen from Fig. \ref{Fig_ConCurrentOverview} that, when multiple independent transmissions appear simultaneously, the receiver can successfully capture the signal of interest (SOI) if its Tx power is sufficiently larger than the sum of interferences.
Capture effect matters on the order of the SOI and interferences.
If the SOI arrives first and its power is higher than the interferences, it is called power capture.
For delay capture, if the SOI arrives some time later, but before the end of preamble symbols of the interfered packets and the SOI's Tx power is sufficiently large, the receiver can also successfully decode the SOI.
Delay capture happens because of radio transceivers's ability to do continuous preamble detection during synchronization process.
In our measurement, when using CC2420 radio transceivers, the maximal temporal displacements for delay capture to function can be $16T_s$ (it is considered $8T_s$ previously), while $T_s=16\mu$s represents one symbol time of IEEE 802.15.4 modulation.
Differing from capture effect, MIM allows a receiver to disengage from an ongoing packet reception, and engage in a new, stronger packet.
An MIM receiver can simultaneously searches for a new (stronger) preamble of the SOI even when locked onto the interferences.
For this reason, MIM reckons on special hardware and upper layer protocol support and is mainly applied in WLAN APs.\\
\indent Recently, Backcast \cite{DuttaSensys10Backcast} and Glossy \cite{ferrari11Glossy} exploit identical packet transmissions to interfere non-destructively.
This physical layer phenomenon is called non-destructive interference (NDI), as illustrated in Fig. \ref{Fig_ConCurrentOverview}.
NDI originates from the scenario that multiple spatially distributed transmitters send an identical packet to a common receiver simultaneously.
Traditionally, we might think concurrent packet transmissions will collide and \emph{prevents} the common receiver from successfully decoding the packet, if Tx power of each transmission is the same (In this case, no capture effect happens).
However, the receiver can decode the packet with high probability if the maximal temporal displacement of concurrent transmissions is within a chip time, namely, 0.5$\mu$s for the IEEE 802.15.4 radio.
Indeed, NDI makes good use of the physical layer tolerance for multi-path signals.\\
\indent \textbf{IEEE 802.15.4 Radio:}
In IEEE 802.15.4, outgoing symbols are mapped to one of 16 pseudo-random, 32-chip sequences by direct-sequence spread spectrum (DSSS) technique.
The duration of each chip is $T_c=0.5\mu$s, while each symbol takes up $T_s=16\mu$s.
The chip sequences are modulated with orthogonal quadrature phase shift keying (O-QPSK) with half-sine pulse shaping.
The demodulator de-spreads 32-chip sequences to 16 valid sequences with smallest Hamming distance.
Symbol synchronization and data decoding are achieved by a continuous preamble and Start of Frame Delimiter (SFD) search.
\subsection{Is NDI constructive?}
\begin{figure*}[t]
  \centering
  \begin{minipage}[b]{0.32\textwidth}
    \centering
    \includegraphics[width=2.45in]{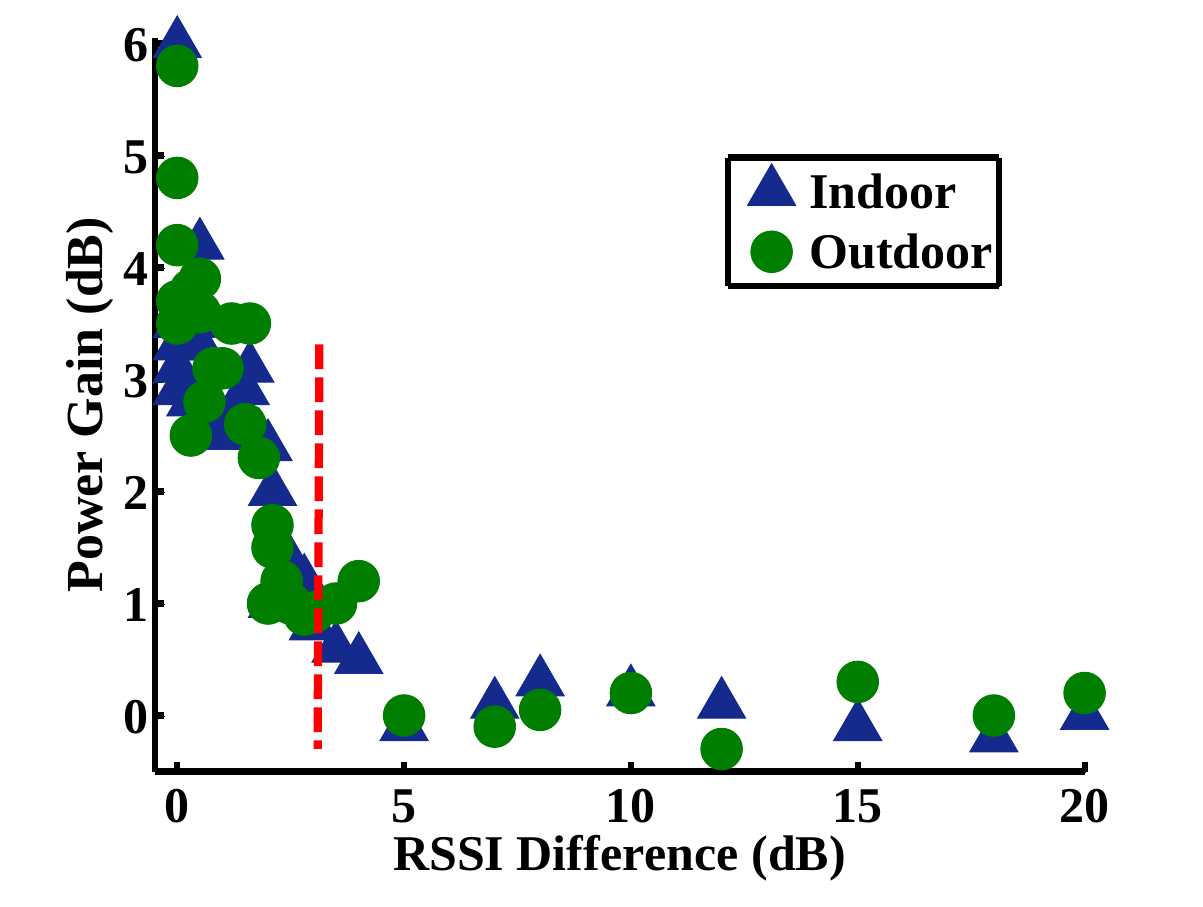}
    \caption{No obvious power gain if the received RSSI differences of R1 and R2 exceed 3dB.}
    \label{Fig_PowerCapture}%
  \end{minipage}%
  \hspace{0.01\textwidth}%
  \begin{minipage}[b]{0.32\textwidth}
    \centering
    \includegraphics[width=2.15in]{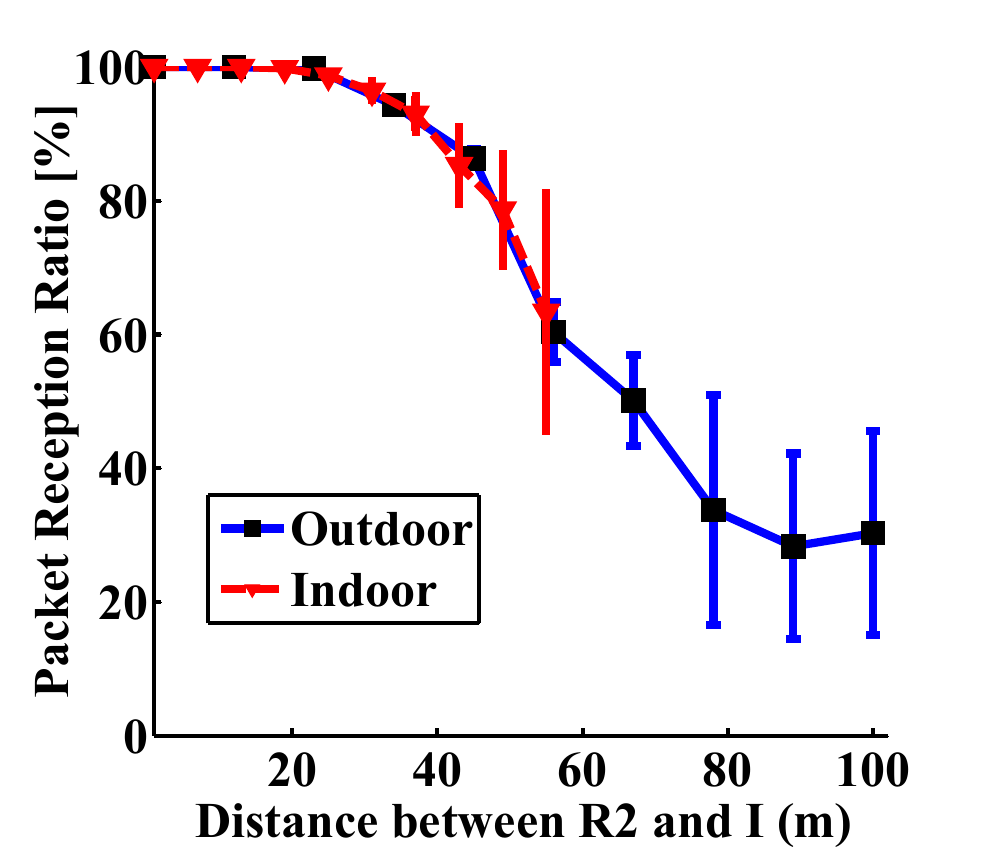}
    \caption{The PRR of NDI drops quickly as the differences of propagation delay increase.}
    \label{Fig_PrapagationDelay}%
  \end{minipage}%
  \hspace{0.01\linewidth}%
  \begin{minipage}[b]{0.32\textwidth}
    \centering
    \includegraphics[width=2.0in]{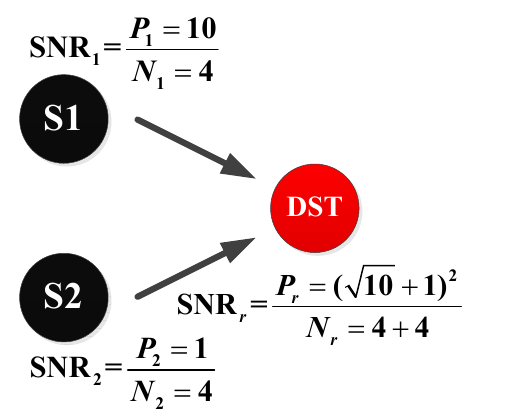}
    \caption{Two signals superpose destructively even if they are perfectly aligned.}
    \label{Fig_LinkSelection}
  \end{minipage}%
 \end{figure*}
\indent The interference due to concurrent packet transmissions is non-destructive if it \emph{doesn't destroy} the normal packet reception.
Since NDI requires different nodes transmitting the same packet and thus may consume more power, more channel resources and require collaboration, one question naturally arises: is NDI constructive?
In other words, is the aggregated effect of multiple concurrent transmissions better than arbitrary single packet transmission?\\
\indent To test whether NDI interfere constructively, we use 3 Tmote Sky sensor nodes, each of which has CC2420 transceiver and MSP430F1611 micro-controller.
One node is selected as the initiator (I), while the other two are chosen as receivers (R1 and R2).
We leverage Glossy source code, an open-source project running on Contiki OS, to generate NDI and
add runtime parameter adaption function.
We do experiments in both outdoor environment and indoor environment.
In each experiment, we fix the positions of I and R1, making them 1 meter away.
We move the position of R2, altering its distance with I from 1 meter to 100 meters (outdoor) or from 1 meter to 55 meters (indoor).
We keep each individual link a perfect link (e.g., PRRs are larger than 99\%), and measure the received PRR and RSSI of I when two receivers transmit packets simultaneously.
The experimental results are illustrated in Fig. \ref{Fig_PowerCapture} and Fig. \ref{Fig_PrapagationDelay}.\\
%
%
\indent To verify whether NDI can give rise to power gain, we fix the positions of R1 and R2 1 meter distance away from I. We measure the received RSSI gain of concurrent transmissions compared with the single better link transmitting individually, drawn as Y-axis in Fig. \ref{Fig_PrapagationDelay}.
The X-axis denotes the received RSSI difference when two receivers transmit independently.
It can be observed from Fig. \ref{Fig_PrapagationDelay} that, the best case of power gain due to two concurrent transmissions can be as high as 6 dB.
However, if the received RSSI difference is larger than 3dB, there is no noticeable power gain.
It is perhaps, \emph{capture effect dominates the packet reception}.
The results indicate that adding more senders \emph{can not} help increase the received RSSI under the condition of capture effect. \\
\indent We also discover that more senders may not lead to PRR improvement. Even worse, more senders might degrade PRR significantly with NDI. We change the distance between R2 and I, and record the PRR when R1 and R2 transmit simultaneously, as shown in Fig. \ref{Fig_PrapagationDelay}. To mitigate the effect of capture effect, we make sure the RSSI values of successful packet receptions of each individual receiver are almost the same. Indeed, due to multi-path effect and external interferences, the received RSSI values are not always stable. We accurately adjust parameters such as Tx powers, antenna directions and retransmission times, to make sure the RSSI values are steady in a short time interval. It can be seen from Fig. \ref{Fig_PrapagationDelay}, when the distance between R2 and I increases, the PRR performance of node I drops distinctly.
The experiments indicate that \emph{propagation delays} also play a crucial role in PRR, even if the differences of transmission distances are only about 40 meters.
In order to ensure concurrent transmissions to interfere constructively, we must compensate propagation delays of spatially distributed transmitters.
\section{A Sufficient Condition For Generating CI}
\label{Sec_SC}
\indent Could we ensure to construct CI, if we compensate different delay uncertainties and perfectly align concurrent transmissions? To answer this question, we first take a simple case in Fig. \ref{Fig_LinkSelection} as an example, and then provide theoretical analysis to provide a sufficient condition for generating CI in WSNs.\\
\indent We suppose sender S1 and S2's signals arriving at the antenna of the destination node have unified signal power 10 and 1, as well as noise power 4 and 4 respectively.
Even if the signals of S1 and S2 exactly align at the common receiver, the effective power of the superposed signal is ${(\sqrt {10}  + 1)^2 { \approx 17.3}}$, while the noise power equals 8.
The SNR (2.17) of superposed signal degrades slightly compared with the single best signal (2.5).
This simple case indicates that only chip level synchronization is not sufficient for CI to function, exactly synchronized signals with different link qualities might also superpose destructively.
In the following discussions, we will focus on \emph{baseband} signals, and examine the role link quality (e.g., PRR, SNR) plays with waveform analysis.\\
\indent The basic principle of 802.15.4 PHY layer is elaborated in \cite{oh2006building}. Let $S_{msk} (t)$ be the transmitted signal after MSK modulation, $I(t)$ and $Q(t)$ denote the
in-phase component and quadrature-phase component respectively.
Let  $\omega _c  =
{\pi  \mathord{\left/
 {\vphantom {\pi  {2T_c }}} \right.
 \kern-\nulldelimiterspace} {2T_c }}
$ represent the angular frequency of half-sine pulse shaping.
The combined MSK signal can be calculated as
\begin{equation}
\label{Equ_MSKsignal}
S_{msk} (t) = I(t)\sin \omega _c t - Q(t)\cos \omega _c t
\end{equation}
\begin{align}
\label{equ_ItQt}
{\rm{where   }}\left\{ \begin{array}{l}
 I(t) = \sum\limits_n {(2C_{2n}  - 1){\rm{rect}}(\frac{t}{2} - nT_c )}  \\
 Q(t) = \sum\limits_n {(2C_{2n + 1}  - 1){\rm{rect}}(\frac{t}{2} - nT_c  - \frac{{T_c }}{2})}  \\
 \end{array} \right.
\end{align}
Here, $C_n  \in \{ 0,1\}$ represents the $n$th chip, and $\rm{rect}()$ function is a rectangle window, defined as
\begin{equation}
\label{Equ_rect}
\rm{rect}(t) \buildrel \Delta \over = \left\{ \begin{array}{l}
 1{\rm{~~~     0}} \le t \le T_c  \\
 0{\rm{~~~     }}\rm{otherwise}. \\
 \end{array} \right.
\end{equation}
After Rayleigh multi-path channel, the received signal $S_R(t)$ is convolution of the original signal and the channel $H(t)$
\begin{align}
\label{Equ_SRt}
 S_R (t) &= S_{msk} (t)*H(t) + N(t).
\end{align}
We suppose there are $N$ transmitters $\{T_i, i=1,2,...,N\}$ simultaneously sending an identical packet to a common receiver $R$.
All the transmissions have already been synchronized at chip level relative to the strongest signal.
In our experiment with CC2420 chip, we find the receivers always synchronize with the strongest signal.
The output signal from each transmitter $T_i$ arriving at the antenna of the receiver $R$ is denoted as $S_R^i(t)$.
Let $\lambda_i$ be SNR of the output signal $S_R^i(t)$, $P_i$ denote average power of signal $S_R^i(t)$ and $N_i$ represent power of noise $N_i(t)$.
Obviously, we have $\lambda_i=\frac{P_i}{N_i}$.
The SNR $\lambda_i$ is mainly determined by the radio propagation environments (e.g., multi-path channels, interferences) and Tx powers of the senders.
The received superposed signal $\overline{S_R (t)}$ is the sum of the $N$ output signals $S_R^i(t)$.
Hence we can approach
\begin{equation}
\label{Equ_CombineRecSignal}
\overline{S_R (t)} = \sum\limits_{i = 1}^N {(A_i S_{R}^i (t - \tau _i )+ N_i(t))},\left| {\tau {}_i} \right| \le T_c
\end{equation}
where $A_i$ and $\tau _i$ respectively depict the unified amplitude and phase offset of the $i$th arriving signal relative to the instant when the strongest signal reaching the receiver. Let $S_{R}^1 (t)$ be the strongest signal. Correspondingly, we have $A_1=1$, $\tau_1=0$, $P_i=P_1{A_i}^2$.
According to \cite{Yin2012SCIF}, it can be derived that the effective power $\overline{P}$ of superposed signals after demodulation is
\begin{equation}
\label{Equ_PowerReceiveSignal}
\overline{P}  =P_1(\sum\limits_{i = 1}^N{A_i \cos } \omega _c \tau _i )^2,
\end{equation}
while the aggregated power of noise $\overline{S_R (t)}$ is
\begin{equation}
\label{Equ_PowerReceiveNoise}
\overline{N}  =\sum\limits_{i = 1}^N{\frac{P_i}{\lambda_i}} .
\end{equation}
As a result, the SNR of the received superposed signal is
\begin{align}
\label{Equ_PowerReceiveNoise}
\frac{\overline{P}}{\overline{N}}= \frac{{P_1 (\sum\limits_{i = 1}^N {A_i \cos \omega _c \tau _i } )^2 }}{{\sum\limits_{i = 1}^N {{{P_i } \mathord{\left/
 {\vphantom {{P_i } {\lambda _i }}} \right.
 \kern-\nulldelimiterspace} {\lambda _i }}} }} \le\frac{{P_1 \sum\limits_{i = 1}^N {A_i^2 } \sum\limits_{i = 1}^N {(\cos \omega _c \tau _i )^2 } }}{{P_1 \sum\limits_{i = 1}^N {{{A_i^2 } \mathord{\left/
 {\vphantom {{A_i^2 } {\lambda _i }}} \right.\kern-\nulldelimiterspace} {\lambda _i }}} }}.
\end{align}
The inequality (\ref{Equ_PowerReceiveNoise}) can be derived by Cauchy-Schwarz inequality and equality holds if the condition satisfies
\begin{align}
\label{Equ_equation}
\frac{{A_i }}{{\cos \omega _c \tau _i }} =\frac{{A_j }}{{\cos \omega _c \tau _j }},~~~~(\forall i,j).
\end{align}
To guarantee the received SNR of superposed signal is better than the SNR of any single signal in the worst case, namely to ensure simultaneous transmissions to interfere positively, it is required that the maximum value of the received SNR is no less than $\lambda _{\max }$
\begin{align}
\label{Equ_minmax}
({\frac{\overline{P}}{\overline{N}}})_{max}>\lambda _{\min } \sum\limits_{i = 1}^N {(\cos \omega _c \tau _i )^2 } \geq \lambda _{\max }.
\end{align}
Consequently, we derive a theoretical \emph{sufficient condition (SC)} for concurrent transmissions with IEEE 802.15.4 radio to interfere constructively.
\begin{enumerate}
\setlength{\itemsep}{-0.245ex}
\item
Concurrent transmissions with an \emph{identical} packet should be synchronized at \emph{chip} level, namely less than $T_c$=0.5$\mu$s;
\item
The phase offset of the $i$th arriving signal should satisfy: $\left| {\tau _i } \right| \le {\raise0.7ex\hbox{${\cos ^{ - 1} \frac{{P_i }}{{P_1 }}}$} \!\mathord{\left/
 {\vphantom {{\cos ^{ - 1} \frac{{P_i }}{{P_1 }}} {\omega _c }}}\right.\kern-\nulldelimiterspace}
\!\lower0.7ex\hbox{${\omega _c }$}}$ (SC-I);
\item
The ratio of the minimum SNR $\lambda _{\min}$ and the maximum SNR $\lambda _{\max}$ of current transmissions should satisfy:
$\frac{{\lambda _{\min } }}{{\lambda _{\max } }} \ge \frac{1}{{\sum\limits_{i = 1}^N {(\cos \omega _c \tau _i )^2 } }}$ (SC-II).
\end{enumerate}
\section{Triggercast Implementation}
\subsection{Triggercast Overview}
\indent In this section, we introduce the implementation of Triggercast to generate CI in WSNs. As illustrated in Fig. \ref{Fig_TriggerCastArchitecture}, Triggercast leverages the instant of a triggered signal as a common reference for all concurrent senders to implement synchronized packet transmissions.
In the MAC layer design, the trigger node utilizes a standard CSMA/CA protocol to acquire the medium.
Once the trigger node senses the channel is free, it first broadcasts a synchronization packet, and tell all the co-senders the destination and when to start forwarding data.
After a promissory duration of time (e.g., tens of ms), all the co-senders begin to transmit simultaneously.\\
\indent In the PHY layer design, Triggercast utilizes our proposed chip level synchronization (CLS) and link selection and alignment (LSA) algorithms to ensure concurrently transmitted packets interfere constructively. For receiver-initiated Triggercast, the receiver first performs LSA to select which links will participate in concurrent transmissions.
For sender-initiated Triggercast, each co-sender individually runs the LSA, to determine whether it will join in the concurrent transmission process.
Then selected senders will use CLS to evaluate propagation and radio processing delays.
Finally, they insert a number of \emph{no operations (NOPs)} (Eq. (\ref{Equ_CompensationNop})) to compensate the evaluated delays and phase offsets obtained in LSA.
It should be noticed that, Triggercast can leverage normal packet transmissions to obtain parameters for CLS and LSA, which can significantly mitigate the overhead.
\subsection{Chip Level Synchronization (CLS)}
\label{Sec_CLS}
\subsubsection{Timing Diagram Analysis}
\label{Sec_TimingAnalysis}
\begin{figure}
\centering
\includegraphics[width=2.65in]{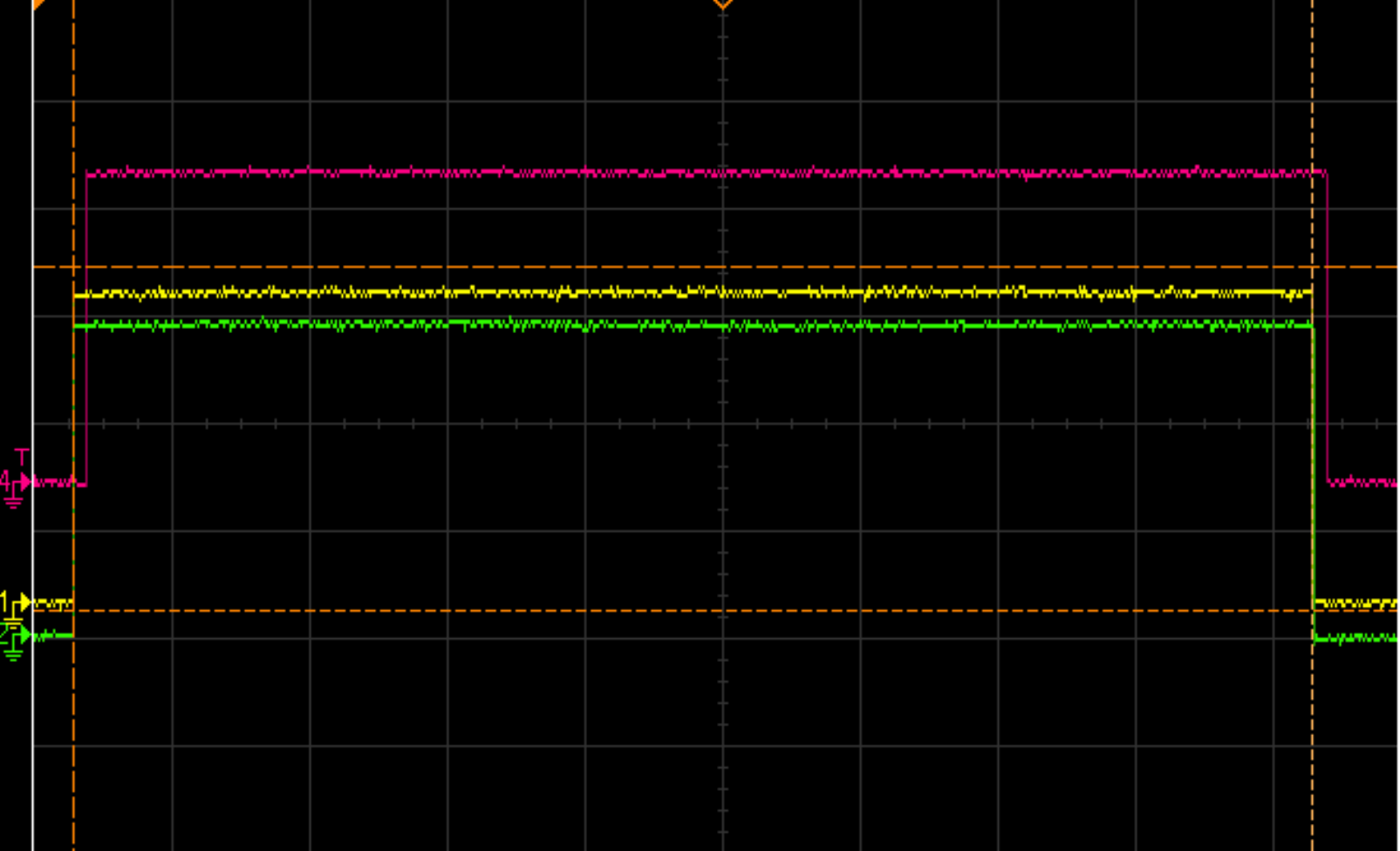}
\caption{SFD pin activities captured by Agilent oscilloscope MSO-X 2024A. Channel 1 and channel 2 display the SFDs of two concurrent transmitted senders. Channel 4 expounds the SFD signal of a receiver.}
\label{Fig_OsillatorSFD}
\end{figure}

\indent In practice, it is very challenging to synchronize outgoing IEEE 802.15.4 symbols at a magnitude of chip level, namely $0.5\mu$s.
One straightforward approach is through accurate time synchronization.
To the best of our knowledge, in WSNs, there is no such time synchronization protocol \cite{MarotiSensys2004FTSP} that can achieve such synchronization accuracy.
Taking the Tmote Sky sensor node as an example, we elaborate fine-grained timing diagram analysis of different delays that might influence the synchronization accuracy of Triggercast.
The Tmote Sky node has an on-board MCU running at a frequency of $f_p=4,194,304$Hz and a CC2420 radio transceiver which updates its digital output with frequency $f_r=8$MHz.\\
\indent  Fig. \ref{Fig_OsillatorSFD} and Fig. \ref{Fig_waveform} illustrate the detailed SFD pin activities during a packet transmission and reception.
From Fig. \ref{Fig_waveform}, it can be noticed that the synchronization accuracy of Triggercast accounts for the propagation delay, the radio processing delay introduced by the radio at the beginning of a packet reception, the hardware turn around delay from the reception state to the transmission state and the software delay. Note that the data transmission delay is a fixed value, determined by packet length. The promissory interval is configured by the MAC layer of Triggercast, and may alter from hundreds of $\mu$s to tens of ms. The uncertainty of promissory interval can be mitigated by classical time synchronization algorithm, which is beyond the scope of this paper.\\
\indent \textbf{(a)Radio processing delay} describes the time between the arrival of the packet at the antenna and the instant when the radio circuits successfully decode the first sample.
Estimating radio processing delay is a challenging task, as it varies from packet to packet, depends on the SNR, as well as the multi-path characteristics of the channel.
Moreover, the asynchronous radio clocks between the transmitter and the receiver also cause a uniform distributed quantization error.\\
\indent \textbf{(b)Propagation delay} is the signal's flight time between transmitter and receiver.
The propagation delay is determined by the distance of a transmitter-receiver pair.\\
\indent \textbf{(c)Software delay} is defined as the duration from the falling edge of the SFD interrupt to the end of a successful packet reception.
The software delay uncertainty mainly depends on variable interrupt serving delays, and the unsynchronized clocks between the MCU and the radio module.
The interrupt serving delays can be accurately evaluated and compensated with the method explained by Glossy \cite{ferrari11Glossy}.
The new generation chip CC2530 integrates MCU and radio module in one chip with synchronized clock frequency, indicating the software delay uncertainty can be perfectly eliminated.\\
\indent \textbf{(d)Hardware turnaround delay} is the time required for a node to switch from packet reception phase to transmission phase.
The hardware turnaround delay is constant, and determined by the speed of the radio frontend from reception to transmission.
\subsubsection{Delay measurement and compensation}
\label{subsec_CLS}
\begin{figure}
\centering
\includegraphics[width=2.8in]{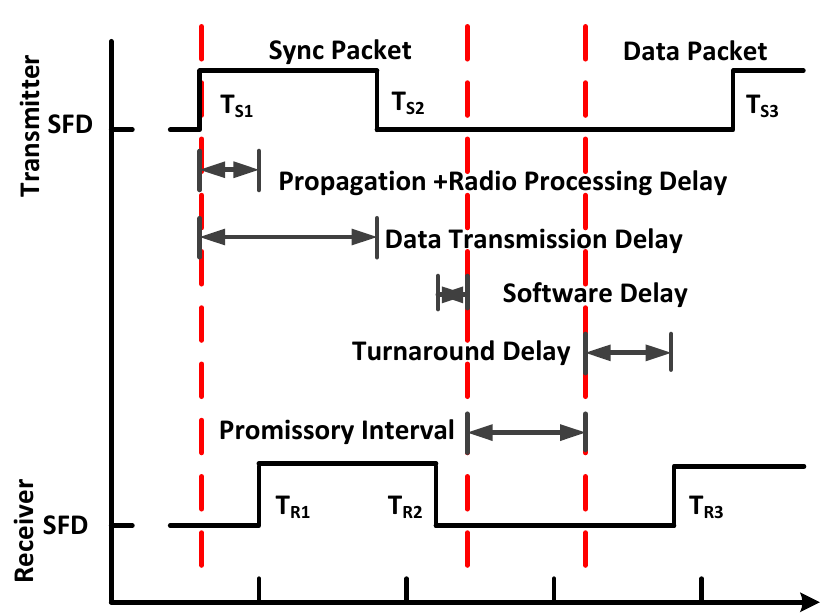}
\caption{The timing diagram of SFD signal for TMote Sky node with Triggercast (Without Preamble).}
\label{Fig_waveform}
\end{figure}
\begin{figure*}[t]
  \centering
  \begin{minipage}[b]{0.32\textwidth}
    \centering
    \includegraphics[width=2.4in]{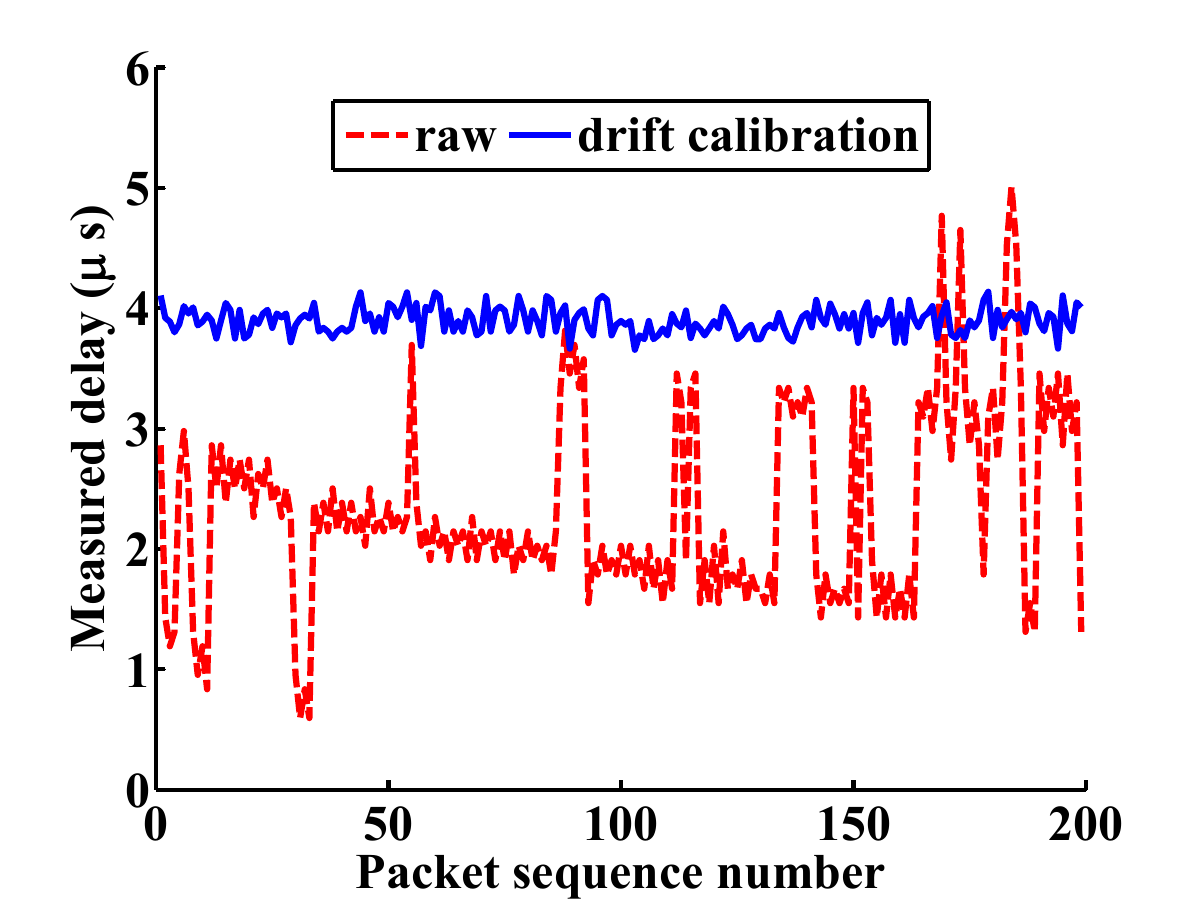}
    \caption{Measured and calibrated delays of propagation and radio processing}
    \label{Fig_DelayCompensation}%
  \end{minipage}%
  \hspace{0.01\textwidth}%
  \begin{minipage}[b]{0.32\textwidth}
    \centering
    \includegraphics[width=2.4in]{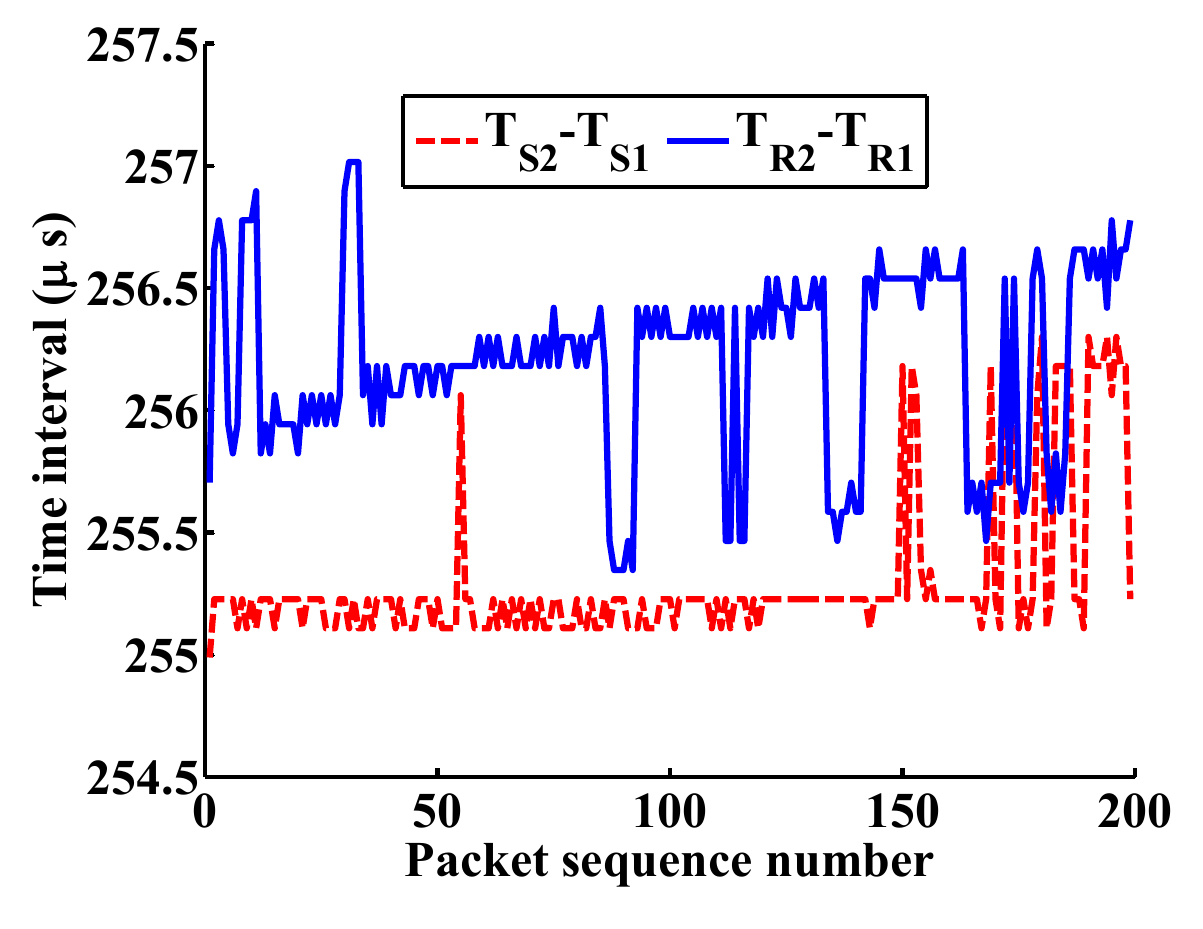}
    \caption{Measured delays of data transmission for the transmitter and the receiver using DCO ticks}
    \label{Fig_DataFlightDiff}%
  \end{minipage}%
  \hspace{0.01\linewidth}%
  \begin{minipage}[b]{0.32\textwidth}
    \centering
    \includegraphics[width=2.4in]{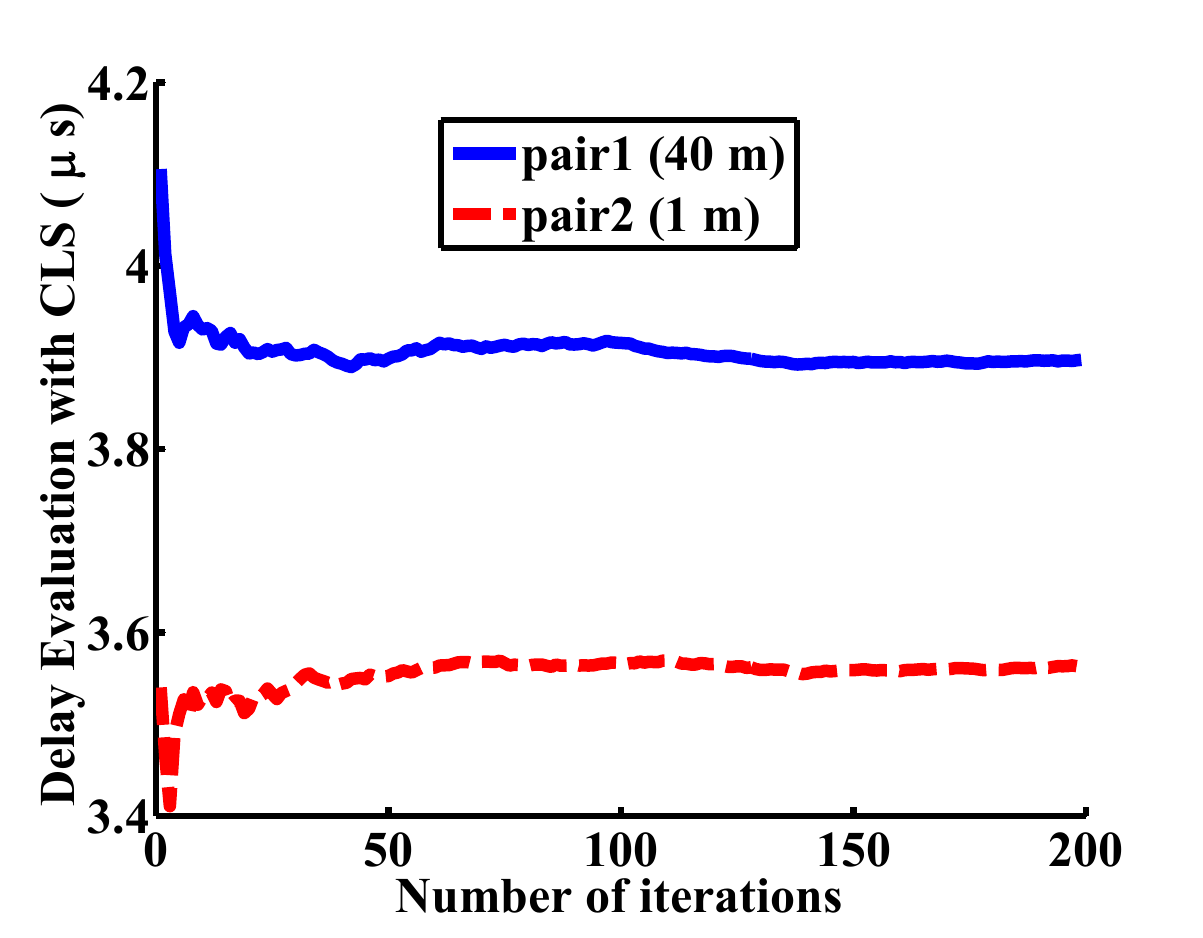}
    \caption{Convergence time analysis of CLS. About 22 runs are enough for delay evaluation.}
    \label{Fig_CoverageTime}
  \end{minipage}%
 \end{figure*}
\indent From Fig. \ref{Fig_PrapagationDelay}, we conclude that propagation delays of spatially distributed transmitters must be compensated, in order to make concurrent transmissions positively superpose.
The TMOTE Sky node has an internal DCO operating at frequency $f_p=4,194,304$Hz, which means the evaluated propagation delays can only have an accuracy of 0.238$\mu$s (about 71.5 meters).
Even worse, the frequency of the DCO can deviate up to $\pm 20\%$ from the nominal value, with temperature and voltage drifts of $-0.38\%^\circ C$ and $5\%/V$.
Moreover, the pair-wise packet transmissions method can only measure the sum of the propagation delay and the radio processing delay.
To subtract the radio processing delay in realistic environment is challenging, as it varies from one packet to another, and is influenced by communication link qualities.
Fortunately, we manifest that the compensation of the \emph{sum} of the propagation delay and the radio processing delay is sufficient for chip level synchronization.
This task is still difficult due to many time uncertainty factors such as the quantization uncertainty, the software delay uncertainty due to asynchronous radio clocks, as well as clock drifts due to packet transmissions.\\
\indent \textbf{Methodology:} According to the \emph{law of large numbers}, the average of the results obtained from a large number of trails should be close to the expected value.
Inspired by this, we select one transmitter-receiver pair which is 40 meters away in indoor environment, and let the transmitter periodically send a packet every 500ms.
Once the receiver successfully decodes a packet, it piggybacks a reply packet as soon as possible to the previous transmitter.
As shown in Fig. \ref{Fig_waveform}, the time-stamps $T_{S1}$ and $T_{S2}$ represent the phases, when the sender's radio starts transmitting a packet and ends a packet transmission, while the time-stamp $T_{S3}$ denotes the phase when the radio begins a packet reception.
The time-stamps $T_{R1}$, $T_{R2}$ and $T_{R3}$ characterize the phases when the receiver's radio starts a packet reception, ends a packet reception as well as begins a packet transmission respectively.
The TMote Sky node can accurately capture the exact instants when MCU detects rising edge and falling edge of SFD interrupts, with MCU's timer capture functionality.
The $n$th packet sent by the receiver includes time-stamps $T_{R1}(n)$, $T_{R2}(n)$ and $T_{R3}(n-1)$, which can be used by the transmitter, to evaluate the expected value of radio processing delay and propagation delay
\begin{align}
\label{Equ_DelayEvaluation}
\widehat{\Delta}=\frac{(\widehat{T_{S3}}-\widehat{T_{S1}})-(\widehat{T_{R3}}-\widehat{T_{R1}})}{2},
\end{align}
where the symbol $\widehat{\lambda}$ defines the mean value of $\lambda$.\\
\indent Experimental results of delay measurement using Eq. (\ref{Equ_DelayEvaluation}) is displayed in Fig. \ref{Fig_DelayCompensation} as the 'raw' curve.
Unfortunately, the result is pessimistic.
The measured delay ranges from 0.596$\mu$s to 5.01$\mu$s, with average value 2.32$\mu$s and variance 0.628$\mu$s.
The instability of measured delay indicates that, it is difficult to synchronize different transmitters at a magnitude of 0.5$\mu$s, if we straightly use the measured data for compensation.
Fortunately, we disclose the data transmission delay is the same for all nodes.
And thus we have
\begin{equation}
\label{Equ_DataTransmissionDelay}
T_{S2}(n)-T_{S1}(n)=T_{R2}(n)-T_{R1}(n).
\end{equation}
The data transmission delays of the transmitter and the receiver are drawn in Fig. \ref{Fig_DataFlightDiff}.\\
\indent We also find that the measured data transmission delays are not stable for the transmitter-receiver pair.
The reason for the instability is because of the jitters, clock drifts as well as hardware diversities of the nodes' DCOs. The drifts can be as high as 5000ppm in our measurement.
We define $\chi (n) = {{(T_{S2} (n) - T_{S1} (n))} \mathord{\left/
 {\vphantom {{(T_{S2} (n) - T_{S1} (n))} {(T_{R2} (n) - T_{R1} (n))}}} \right.
 \kern-\nulldelimiterspace} {(T_{R2} (n) - T_{R1} (n))}}$ as the unified clock drift coefficient relative to the receiver.
Consequently, we can calibrate Eq. (\ref{Equ_DelayEvaluation}) as
\begin{equation}
\label{Equ_CalibrationDelay}
\widehat{\Delta}_{cal}=\frac{\text{mean}(\frac{T_{S3}(n)-T_{S1}(n)}{\chi (n)})-(\widehat{T_{R3}}-\widehat{T_{R1}})}{2}.
\end{equation}
We obtain the expected radio processing and propagation delay represented by DCO Ticks after the calibration of Eq. (\ref{Equ_CalibrationDelay}).
To translate them to time, we also utilize the \emph{Virtual High-resolution Time (VHT)} \cite{SchmidIPSN10} approach, which calibrates the receiver's DCO with more stable external 32,768 Hz crystal as a reference.
The measured propagation and radio precessing delay after clock drift calibration is shown as the 'drift calibration' curve in Fig. \ref{Fig_DelayCompensation}.
The calibrated delay ranges from 3.66$\mu$s to 4.12$\mu$s, with average value 3.90$\mu$s and variance 0.012$\mu$s.\\
\indent \textbf{Convergence Time:} To measure CLS's convergence performance, we do experiments with two transmitter-receiver pairs.
Pair 1 and pair 2 are 40 meters and 20 meters away respectively.
We average the calibrated delay with Eq. (\ref{Equ_CalibrationDelay}), and test how well CLS works in terms of convergence time.
Experimental results are shown in Fig. \ref{Fig_CoverageTime}.
The medium convergence time are 22 runs.
Since each run can be done in at least 892 $\mu$s, the delay evaluation algorithm consumes about 20ms and can be accomplished adaptively as the channel state dramatically changes.
We disclose that, in our measurements, the delays don't change so much as thought before.
The measurement delay are almost constant, unless the nodes move or the channel significantly changes.
%
\subsection{Link Selection and Alignment (LSA)}
\begin{algorithm}
\label{Alg_LSA}
\caption{Link Selection and Alignment} \SetKwData{Index}{Index}
\SetAlgoNoLine  
\LinesNumbered
\KwIn{Given a lossy link set $\Phi<P_i,\lambda_i>$, where $P_i$ and $\lambda_i$ represent the received RSSI and SNR. All link pairs of $\Phi$ are ordered} \KwOut{A lossy link subset $\Omega<P_j,\lambda_j,\tau_j>$ to maximize the superposed signal's SNR, where $\tau_j$ is the maximal allowed phase offset.}
\BlankLine
Sort $\Phi$ with arbitrary optimal sorting algorithm, and store the result as $\Phi'$ \\
Get the best link $<P,\lambda>$ in $\Phi'$
Insert link $<P,\lambda>$ and "0" (phase offset) in empty set $\Omega$\\
\For{$i=2:N$}
{{get the best link $<P_i,\lambda_i>$ in sorted set $\Phi'$, store as link $<P_t,\lambda_t>$}\;
{calculate maximal allowed phase offset $\tau_i$ of link $<P_t,\lambda_t>$ using link $<P,\lambda>$, to satisfy (SC-I) of the sufficient condition provided in Section \ref{Sec_SC}}\;
{use set $\Omega$, link $<P_t,\lambda_t>$, phase offset $\tau_i$ to verify SC-II of the sufficient condition}\;
\If {SNR of link $<P_t,\lambda_t>$ satisfies SC-II}
{{insert link $<P,\lambda>$, phase offset $\tau_i$ to set $\Omega$}\;
\Else{\textbf{break;}}}}
\end{algorithm}
\begin{figure*}[t]
  \centering
  \begin{minipage}[b]{0.32\textwidth}
    \centering
    \includegraphics[width=1.9in]{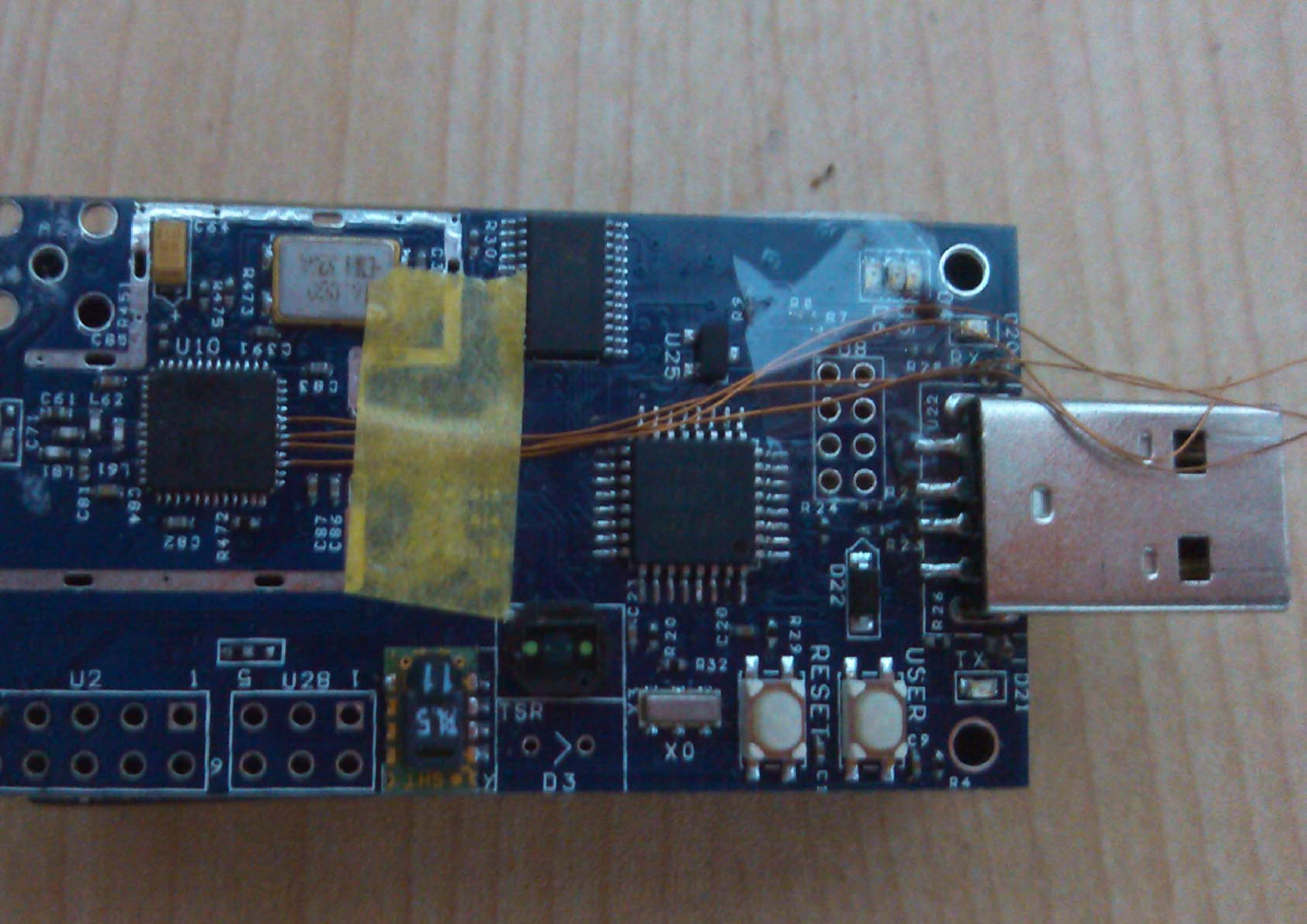}
    \caption{We probe pin activities from CC2420 with very thin enameled wire.}
    \label{Fig_testbed_node}%
  \end{minipage}%
  \hspace{0.01\textwidth}%
  \begin{minipage}[b]{0.32\textwidth}
    \centering
    \includegraphics[width=1.9in]{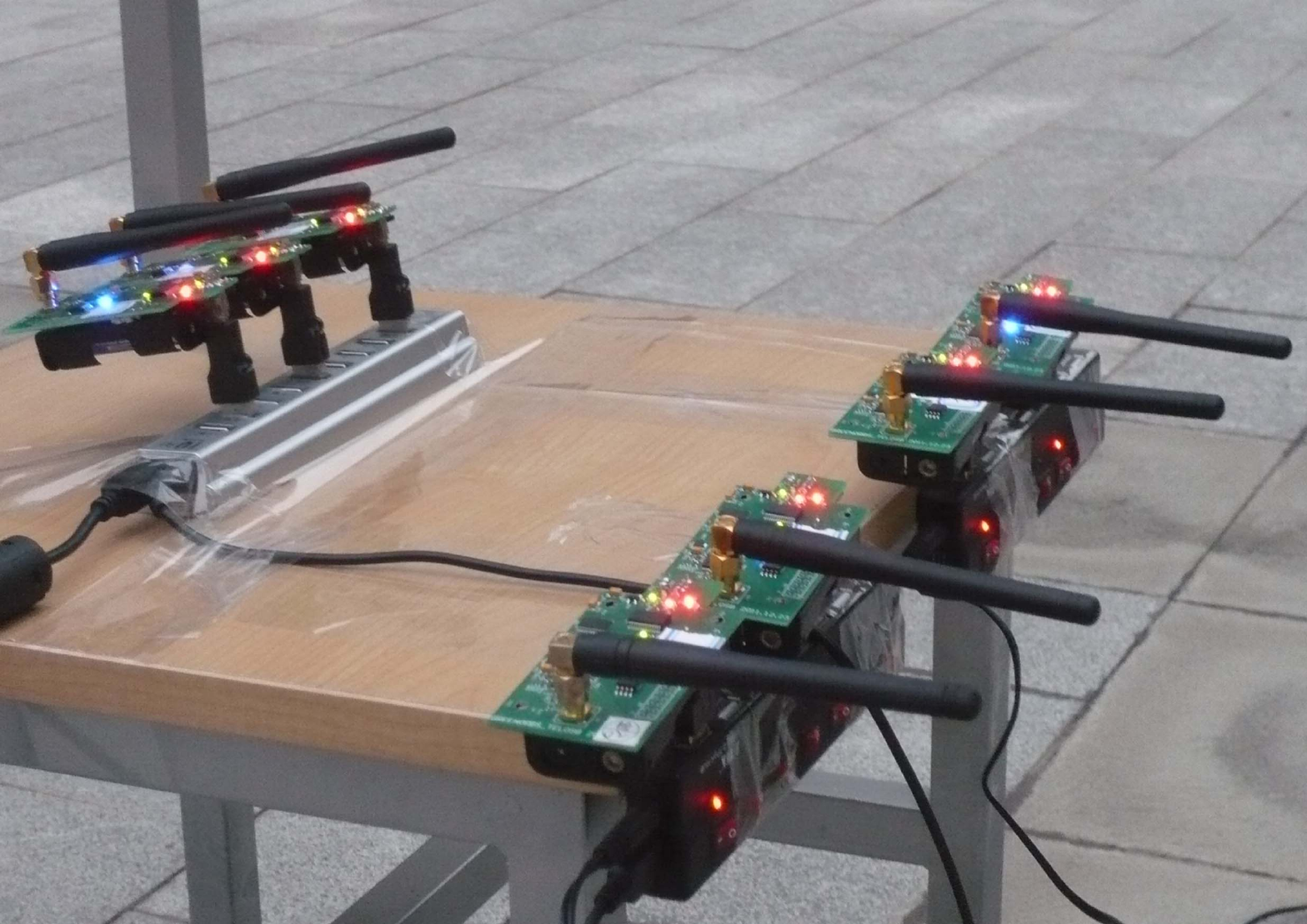}
    \caption{A snapshot of part of our testbed.}
    \label{Fig_Testbed}%
  \end{minipage}%
  \hspace{0.01\textwidth}%
  \begin{minipage}[b]{0.32\textwidth}
    \centering
    \includegraphics[width=2.11in]{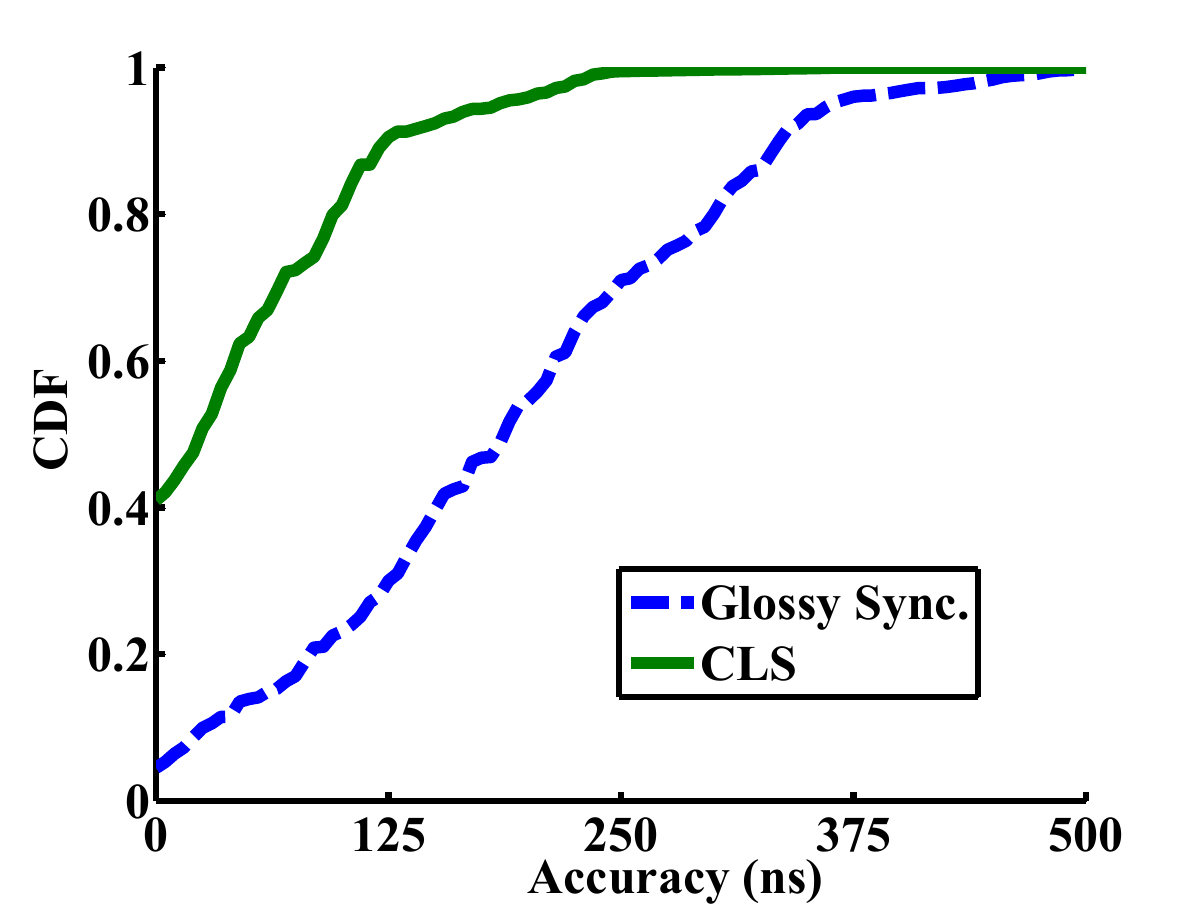}
    \caption{Synchronization error of Triggercast less than 250 ns has more than 95\% confidence}
    \label{Fig_SynchronizationAccuracy}%
  \end{minipage}%
\end{figure*}
\indent Note that SNR can be derived from PRR through theoretical models \cite{Yin2012SCIF} or online measurements \cite{Xing2010ICNPPassive}.
The relationships between SNR and PRR can be mapped as look-up tables and stored in the external flashes of sensor nodes for Triggercast to use. Assuming all the concurrent transmissions are synchronized at the chip level with CLS, according to the proposed sufficient condition in Section \ref{Sec_SC}, the problem to make concurrent transmissions superpose constructively can be formalized as CI-generation problem.\\
\indent \emph{Problem:} Let $\Phi = \{(P_1,\lambda_1),(P_2,\lambda_2),...,(P_N,\lambda_N),\}$ define a lossy link set, where $P_i$ and $\lambda_i$ denote the received signal's RSSI and SNR of transmitter $T_i$ respectively.
The problem is to find a lossy link subset $\Omega$, in order to maximize the superposed signal's SNR on condition that the combined link is better than any lossy link in $\Phi$ and the phase offset $\tau_i$ is as large as possible.\\
\indent We define a link pair $(L_i, L_j)$ is \emph{ordered} if $P_i\geq P_j$ indicates $\lambda_i\geq \lambda_j$, $1 \leq i,j \leq N$.
According to the sufficient condition for CI, it can be proved that this problem is NP-hard if there exists disordered link pair in $\Phi$.
In practice, it is reasonable to assume all the link pairs in $\Phi$ are ordered.
Based on this assumption, we will show this problem can be solved in $O(nlog~n)$ time. \\
\indent The pseudocode of LSA is described in algorithm \ref{Alg_LSA}. The for-loop can safely break if $\lambda_i$ doesn't satisfy SC-II of the sufficient condition.
For all $\lambda_j \leq \lambda_i$, we have $\tau_j \geq \tau_i$ and thus we can prove that they all don't satisfy SC-II.
\begin{align}
\label{Equ_SafeBreak}
\frac{{\lambda _j }}{{\lambda _{\max } }} \le \frac{{\lambda _i }}{{\lambda _{\max } }} &= \frac{1}{{\sum\limits_{k = 1}^{i - 1} {(\cos \omega _c \tau _k )^2  + (\cos \omega _c \tau _i )^2 } }}\nonumber\\
&\le \frac{1}{{\sum\limits_{k = 1}^{i - 1} {(\cos \omega _c \tau _k )^2 }  + (\cos \omega _c \tau _j )^2 }}
\end{align}
\indent \textbf{Time Complexity:} the time complexity of LSA algorithm is dominated by the sort function.
Clearly, the optimal sort algorithm has an $O(nlog~n)$ time complexity.
Thus the time complexity of LSA algorithm is also $O(nlog~n)$.\\
\indent \textbf{Total compensation time:} we let $\tau _i$ as large as possible to obtain the system's maximal robustness for synchronization errors. Consequently, the total number $N_{com}$ of NOPs for co-sender $T_i$ in Triggercast is
\begin{align}
\label{Equ_CompensationNop}
N_{com}=\left[ {(T - \widehat\Delta _{cal}  + \tau _i )f_p } \right]
\end{align}
where $[]$ is the round function, and $T$ is a predefined maximum delay calibration time.
\section{Performance}
\label{Sec_experiments}
We have implemented a prototype Triggercast on TMote sky sensor nodes.
The software is based on Contiki OS.
During the overall Triggercast's duration, except for the promissory interval, all the relevant interrupts and hardware timers that are not essential to Triggercast's functioning are disabled.
Since this interval is very short (several milliseconds), it is feasible that Triggercast doesn't influence the upper layer's functionality.
A runtime parameter adjustment software is developed, to make sure we can online change the system running parameters, without altering communication channels by programming the nodes.
We test the performance of Triggercast in a practical testbed (Fig. \ref{Fig_Testbed}).
\subsection{Synchronization Accuracy}
\label{Subsec_SychronizationAccuracy}
\begin{figure*}[t]
  \centering
  \begin{minipage}[b]{0.32\textwidth}
    \centering
    \includegraphics[width=2.4in]{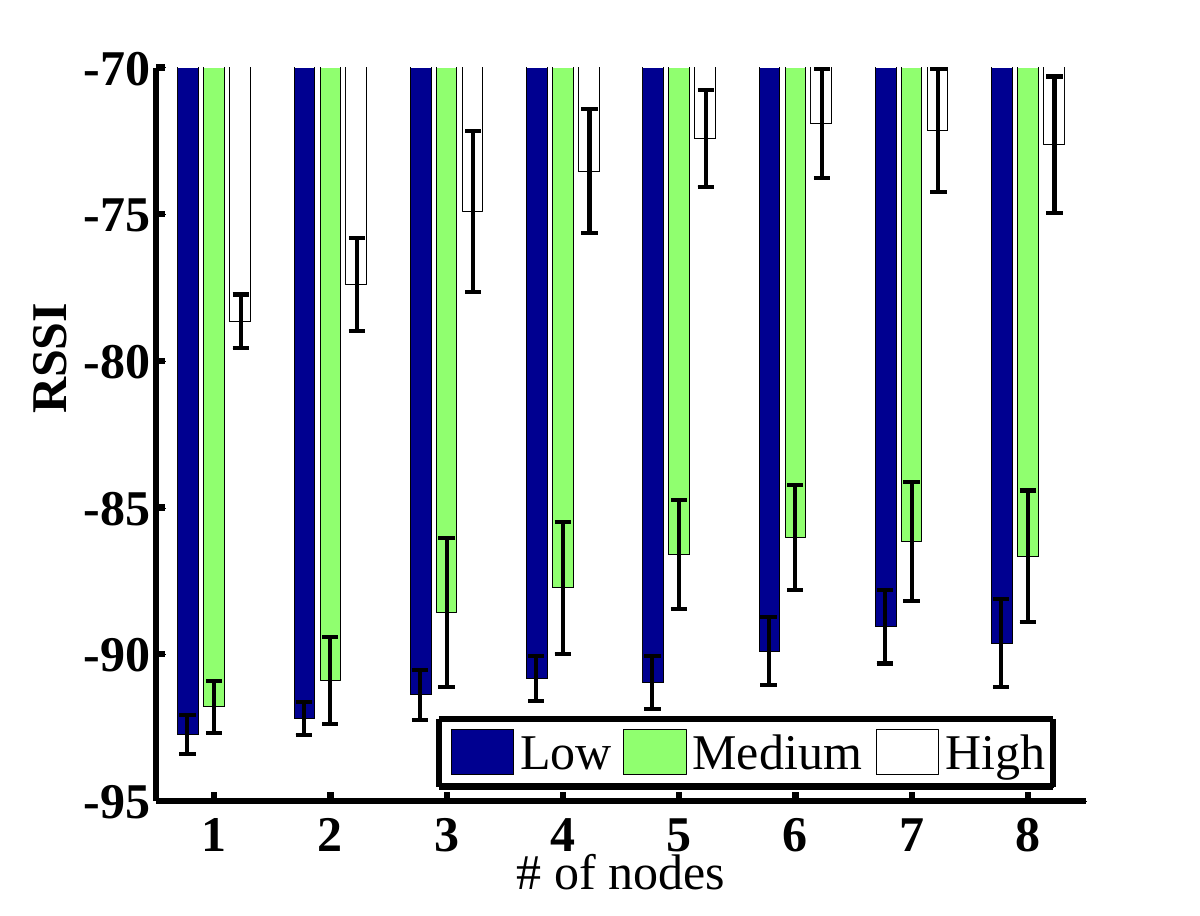}
    \caption{Triggercast increases RSSI.}
    \label{Fig_CIvsRSSI}%
  \end{minipage}%
  \hspace{0.01\linewidth}%
  \begin{minipage}[b]{0.32\textwidth}
    \centering
    \includegraphics[width=2.4in]{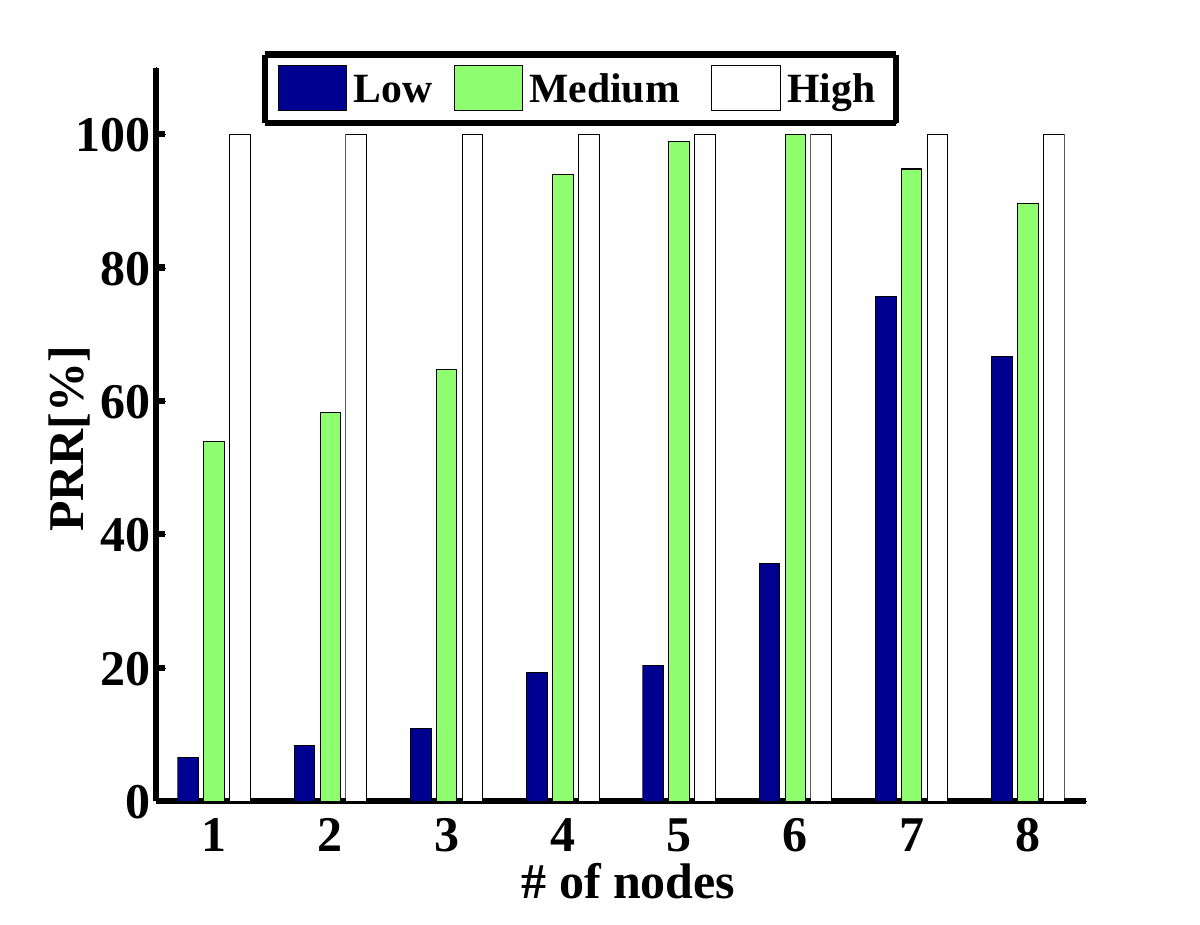}
    \caption{Triggercast improves PRR.}
    \label{Fig_CIvsPRR}
  \end{minipage}%
  \hspace{0.01\linewidth}%
  \begin{minipage}[b]{0.32\textwidth}
    \centering
    \includegraphics[width=2.4in]{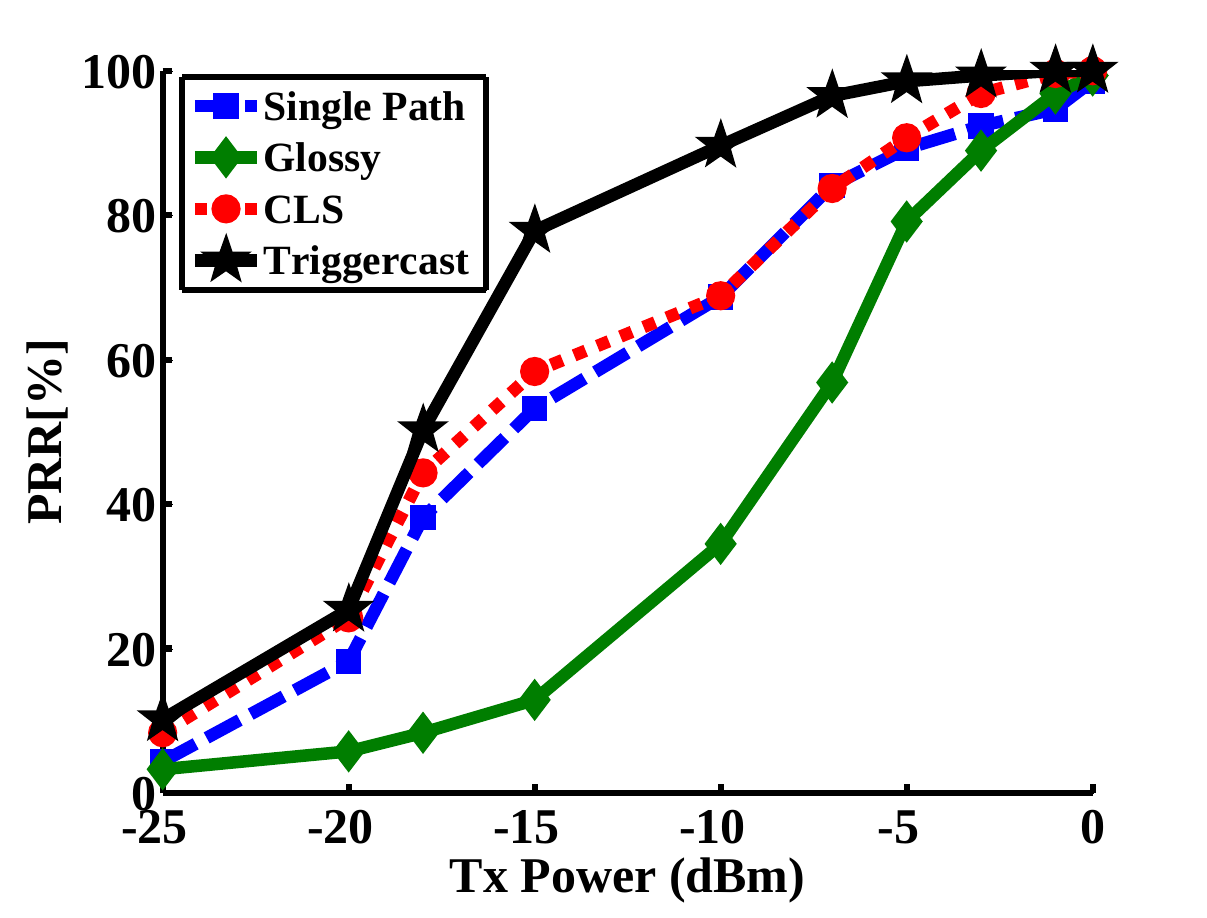}
    \caption{Triggercast provides PRR gains over traditional single-path routing.}
    \label{Fig_DataForwardingExperiments}
  \end{minipage}%
 \end{figure*}
\indent We first test the synchronization performance of multiple concurrent transmitters.
We use three TMote sky nodes, one as a receiver and two as transmitters.
We set the promissory interval parameter to 0 in receiver-initiated Triggercast.
We connect the SFD pins of the receiver (R) and one of the transmitters (S1) to a Agilent MSO-X serial oscilloscope (Fig. \ref{Fig_testbed_node}).
The other transmitter (S2) is 30 meters away in an indoor environment.
However, it is difficult to measure the synchronization of S1 and S2 directly with the oscilloscope.
As a result, we use R as a reference node.
The synchronization between S1 and R can be monitored by the oscilloscope with a granularity of 5ns.
The durations between $T_{R1}$ and $T_{R3}$ (Fig. \ref{Fig_waveform}) of the receiver are accurately measured when S1 and S2 transmit independently.
The differences of the durations can be used for synchronization accuracy measurement, since both S1 and S2 rely on the instant $T_{R1}$ as a reference.
The CDF of synchronization errors compared with the Glossy synchronization algorithm is illustrated in Fig. \ref{Fig_SynchronizationAccuracy}.
Triggercast's CLS algorithm can synchronize multiple transmitters at a magnitude of 250ns.
The accuracy is limited by the operating frequency of the MCU of TMote Sky sensor nodes.
The Glossy synchronization algorithm degrades as the distance differences between two transmitter-receiver pairs increase.
CLS outperforms Glossy because CLS compensates the time due to propagation and radio processing delays.
\subsection{Power Gains and PRR Improvements}
\label{subsec_NEWLINK}
\indent The main purpose of this paper is to make wireless collisions interfere constructively.
In other words, multiple senders transmit an identical packet simultaneously can improve RSSI and PRR.
We do experiments by carrying out Triggercast in both indoor and outdoor environments with our testbed (Fig. \ref{Fig_Testbed}).
Up to 8 senders with all three different kinds of links (low (PRR $<$ 5$\%$), medium (5$\%<$PRR$<90\%$), high(PRR $>$ 90$\%$)), are executed to transmit packets at the same time.
Due to the limit of physical space, we randomly insert NOPs to simulate different propagation and radio processing delays.
We adjust the received RSSIs of each sender's individual packet transmission to almost the same, to eliminate the influence of capture effect.
All the results are averages of more than 1000 tests.
Fig. \ref{Fig_CIvsRSSI} lists the power gains due to multiple senders of different link types (disconnected link: 1-5 dB, intermediate link: 2-6 dB, connected link: 2-6 dB).
The maximum power gain can approach $N^2$ for $N$ concurrent transmitters.
Fig. \ref{Fig_CIvsPRR} shows that PRR can be significantly improved by leveraging CI.
For 7 disconnected links, the PRR achieves almost 70\%, which is better than our previous understandings of harnessing sender diversity gain ($1-(1-0.05)^3\approx 30.2\%$).
Triggercast improves the PRR of intermediate links from 50\% to almost 100\% with 6 concurrent senders.
Our experiments indicate that Triggercast can \emph{control network topology} (e.g., increasing new communication links) \emph{without changing the original network state} (adding new nodes, increasing nodes' power, etc.).
This characteristic is attractive to improve routing performance (explained in Fig. \ref{Fig_OpportunisticRouting}).
To the best of our knowledge, we are the \emph{first} to report multiple concurrent transmitters can reach such PRR improvements in realistic WSNs.
\subsection{Data forwarding with Triggercast}
\indent We create a five node topology as in Fig. \ref{Fig_OpportunisticRouting}(b).
We select two nodes as the source node and the receiver node respectively.
The other three nodes perform forwarding.
All five nodes are placed in random positions in our office.
One of the three relay nodes is deployed near the window, exposure to sunshine.
We also use a hair dryer to heat the node to increase the DCO jitters and decrease its link quality.
We first measure pairwise loss rates between the nodes to compute the ETX metric for each link.
We also evaluate the propagation and radio processing delays with CLS algorithm.
We fix Tx power of source node as 0dBm, online adjust Tx power of relay nodes from -25dBm to 0dBm, and record the PRR of the receiver.
Experimental results are elaborated in Fig. \ref{Fig_DataForwardingExperiments}.\\
\indent As expected, exploring CLS and LSA together results $1.3\times$ PRR gains on average over traditional single-path routing.
The gain of using CLS alone is not obvious.
The reason is because the dirty link (the heated node/receiver pair) influences the overall performance.
Glossy works even worse than single-path routing. The reason is beacuse Glossy only generates NDI, and doesn't compensate propagation and radio processing delays, as well as make link selections.
\section{Conclusions and future work}
\label{Sec_Conclusion}
\indent We introduce Triggercast, the \emph{first} work to implement CI instead of NDI in WSNs, to the best of our knowledge.
Triggercast compensates propagation and radio processing delays, and makes link selection as well as transmission alignment, in order to construct CI.
We implement Triggercast in real testbed, and experimentally demonstrate that Triggercast produces significant performance gains in data forwarding protocols.
We also provide a theoretical sufficient condition on how to ensure concurrent transmissions interfere constructively.
Future work includes adding node mobility and low duty-cycle factors in Triggercast, and exploiting Triggercast in time synchronization and localization.
We are also developing Triggercast as an independent service and open source software component to the community.


\ifCLASSOPTIONcaptionsoff
  \newpage
\fi



%
\bibliographystyle{IEEEtran} 
\bibliography{IEEEabrv,wyWSNRef}

\end{document}